\documentclass[12pt]{article}

\usepackage{orcidlink,lmodern}
\usepackage{graphicx, float, caption}
\usepackage{natbib}
\usepackage{authblk} 
\usepackage{multirow}
\usepackage{amsmath, amsfonts, amssymb}
\usepackage{hyperref}

\definecolor{orange1}{RGB}{255,128,0}
\definecolor{purple2}{RGB}{102,0,204}
\definecolor{blue}{RGB}{0,0,255}
\definecolor{red}{RGB}{255,0,0}
\definecolor{grey}{RGB}{192,192,192}

\newcommand{\MC}{\text{MC}}
\newcommand{\IQR}{\text{IQR}}
\newcommand{\med}{\text{med}}
\newcommand{\MAD}{\text{MAD}}

\newcommand{\cov}{\text{cov}}
\newcommand{\MD}{\text{MD}}
\newcommand{\RD}{\text{RD}}
\newcommand{\bdot}{\boldsymbol{\cdot}}

\newcommand{\ba}{\boldsymbol{a}}
\newcommand{\bx}{\boldsymbol{x}}
\newcommand{\bbeta}{\boldsymbol{\beta}}
\newcommand{\hbbeta}{\boldsymbol{\hat \beta}}
\newcommand{\bmu}{\boldsymbol{\mu}}
\newcommand{\hmu}{\hat{\mu}}
\newcommand{\bhmu}{\hat{\bmu}}
\newcommand{\hsigma}{\hat{\sigma}}
\newcommand{\bSigma}{\boldsymbol{\Sigma}}

\newcommand{\pkg}[1]{{\normalfont\fontseries{b}\selectfont #1}}

\let\proglang=\textsf 
\let\code=\texttt

\begin{document}

\def\spacingset#1{\renewcommand{\baselinestretch}
{#1}\small\normalsize} \spacingset{1}

\title{\textbf{RobPy}: a \textsf{Python} Package for Robust Statistical Methods}

\author[1]{Sarah Leyder}
\author[1]{Jakob Raymaekers}
\author[2]{Peter J. Rousseeuw}
\author[1]{\\Thomas Servotte}
\author[1]{Tim Verdonck}

\affil[1]{Department of Mathematics, University 
          of Antwerp, Belgium}
\affil[2]{Section of Statistics and Data Science, 
          University of Leuven, Belgium}

\setcounter{Maxaffil}{0}
\renewcommand\Affilfont{\itshape\small}
\date{November 2, 2024}           
  \maketitle

\bigskip
\begin{abstract}
Robust estimation provides essential tools for analyzing data that contain outliers, \mbox{ensuring} that statistical models remain reliable even in the presence of some anomalous data. While robust methods have long been available in \proglang{R}, \proglang{Python} users have lacked a comprehensive package that offers these methods in a cohesive framework. \pkg{RobPy} addresses this gap by offering a wide range of robust methods in \proglang{Python}, built upon established libraries including \pkg{NumPy}, \pkg{SciPy}, and \pkg{scikit-learn}. This package includes tools for robust preprocessing, univariate estimation, covariance matrices, regression, and principal component analysis, which are able to detect outliers and to mitigate their effect. In addition, \pkg{RobPy} provides specialized diagnostic plots for visualizing outliers in both casewise and cellwise contexts. This paper presents the structure of the \pkg{RobPy} package, demonstrates its functionality through examples, and compares its features to existing implementations in other statistical software. By bringing robust methods to \proglang{Python}, \pkg{RobPy} enables more users to  perform robust data analysis in a modern and versatile programming language.
\end{abstract}


\section[Introduction]{Introduction} \label{sec:intro}

Data containing outliers pose severe challenges to data scientists. Outliers can heavily distort the outcome of a statistical analysis, and should thus be handled carefully. Robust statistics develops a collection of tools that are designed to provide reliable results even when outliers are present in the data. The approach taken by robust methodology is to first fit a model to the clean part of the data, and then to detect potential outliers through their deviation from this robust fit \citep{rousseeuw1987bookdata,maronna2019robust}.

Over the last two decades, \proglang{R} has been the dominant programming language for implementing and disseminating robust statistical methods. Packages such as \pkg{Robustbase} \citep{robustbase} and \pkg{rrcov} \citep{rrcov} are widely used for robust statistical analysis, with \pkg{Robustbase} alone having nearly 10 million downloads. More specialized packages like \pkg{robustHD} \citep{robustHD} for high-dimensional data, and \pkg{cellWise} \citep{cellWise} focused on cellwise outliers, have further extended robust statistical methodologies in \proglang{R}. Additionally, some of these algorithms have been implemented in \mbox{\proglang{MATLAB}} through the \pkg{LIBRA} library \citep{LIBRA} and the \pkg{FSDA} library \citep{riani2012fsda}.

Despite the growing popularity of \proglang{Python} in data science, the availability of robust statistical methods in \proglang{Python} has remained limited. Aside from a few implementations, such as the \code{MinCovDet} function
in the \pkg{scikit-learn} library \citep{scikit-learn} and the \code{RLM} function for regression M-estimators in \pkg{statsmodels} \citep{seabold2010statsmodels}, \proglang{Python} users have had little access to the robust tools readily available in \proglang{R}.

To address this gap we created \pkg{RobPy}, a \proglang{Python} package that consolidates the most popular robust statistical algorithms under one umbrella. Built on established libraries like \pkg{NumPy}, \pkg{SciPy}, and \pkg{scikit-learn}, \pkg{RobPy} provides robust tools for data preprocessing, univariate estimation, covariance matrices, regression, and principal component analysis. These tools are complemented by specialized visualization techniques for diagnosing and handling outliers. \pkg{RobPy} aims to bring robust data science algorithms to a wider audience, enabling \proglang{Python} users to perform reliable data analysis even in the presence of outliers.

We will first describe the structure of the \pkg{RobPy} package and explain how it builds on established existing \proglang{Python} libraries for statistics and data science. In Section \ref{sec:softwareusage} we then provide practical examples of the package's functions applied to real-world data.

\section{Structure} \label{sec:structure}

\pkg{RobPy} is a \proglang{Python} package specifically created for robust statistical analysis. It offers an extensive array of modules, classes and functions. The package aims to be user-friendly, while providing a broad spectrum of functionalities. \pkg{RobPy} inherits from the popular libraries \pkg{NumPy} \citep{numpy}, \pkg{SciPy} \citep{SciPy} and \pkg{scikit-learn} \citep{scikit-learn}, ensuring compatibility with these foundational tools. Many of its algorithms mirror or are inspired by implementations found in the \proglang{R}-packages \pkg{robustbase}, \pkg{rrcov} and \pkg{cellWise}. 

In the \pkg{RobPy} package, most base classes are derived from the \pkg{scikit-learn} API \citep{scikit-learn_api}. By using their base classes and adhering to their conventions our implementations are standardized, making them more familiar for users. This also allows users to seamlessly integrate our algorithms with other \pkg{scikit-learn} tools. All of the algorithms are implemented in an object-oriented way and adhere to the \code{fit-predict}/\code{fit-transform} conventions used by \pkg{scikit-learn}.

Most \pkg{RobPy} classes include specialized tools to visualize outliers. In robust statistics, many methods have diagnostic plots to help distinguish between normal observations and various types of outliers. Consequently, we equipped most base classes with a tailored plotting tool specific to the analysis method used. This will be illustrated in Section \ref{sec:softwareusage}.

In the following subsections we will discuss the structure of the various \pkg{RobPy} modules, detailing the classes and functions they implement. We will reference the papers that discuss the robust methods used and their original implementations. Table \ref{tab:allmethods} provides a quick overview of all algorithms implemented in each module and their corresponding visualization tools. For a detailed description of the individual functions and their parameters we refer to the \pkg{RobPy} documentation which can be found at \href{https://robpy.readthedocs.io/en/latest/index.html}{robpy.readthedocs.io}\,.

\begin{center}
\begin{table}[H] \small
\setlength{\tabcolsep}{3pt}
\begin{tabular}{|l||l|l|l|}
\hline
Module & Classes and/or methods & Description & Visualization \\
\hline
\hline
\multirow{3}{*}{Preprocessing} & \code{DataCleaner} & cleans a dataset for analysis & \multirow{3}{*}{} \\
         & \code{RobustPowerTransformer} & robustly transforms features & \\
        & \code{RobustScaler} & robustly scales features &  \\
\cline{1-3}
\multirow{5}{*}{Univariate} & \code{UnivariateMCD} & univariate MCD estimator &  \\
                  & \code{OneStepM} & one-step M-estimator &                   \\
                  & \code{Qn} & $Q_n$ estimator of scale &                   \\
                  & \code{Tau} & $\tau$ estimator of scale&                   \\\cline{4-4}
                  & \code{adjusted\_boxplot} & boxplot for skewed data &  returns the boxplot \\
\hline
\multirow{6}{*}{Covariance} & \code{FastMCD} & fast MCD algorithm &  \multirow{5}{3.7cm}{distance-distance:\\ robust distances versus \\ Mahalanobis distances\\} \\
                 & \code{DetMCD} & deterministic MCD algorithm &  \\
                 & \code{WrappingCovariance} & covariance using wrapping & \\
                 & \code{KendallTau} & covariance via Kendall's $\tau$ &                  \\
                 & \code{OGK} & OGK covariance matrix & \\\cline{4-4}
                 & \code{CellMCD} & cellwise robust MCD & five diagnostic plots                \\ 
\hline
\multirow{3}{*}{Regression} & \code{FastLTSRegression}& fast Least Trimmed Squares & \multirow{3}{3.7cm}{plot of robust\\ residuals versus \\ robust distances of $x$ \\ } \\
 & \code{SRegression} & fast S-regression & \\
 & \code{MMRegression} & MM-regression & \\
 \hline
 \multirow{2}{*}{PCA} & \code{ROBPCA} & ROBust PCA & \multirow{2}{3.7cm}{plot of orthogonal\\
  versus score distances}\\
  & \code{PCALocantore} & Spherical PCA & \\
  \hline
  Outliers & \code{DDC} & detects deviating cells & cellmap \\
  \hline
  \multirow{6}{*}{Utils} & \code{mahalanobis\_distance} & Mahalanobis distances & \multirow{6}{*}{}\\
  & \code{l1median} & spatial median for location & \\
  & \code{weighted\_median} & weighted univariate median & \\
  & \code{stahel\_donoho} & Stahel-Donoho outlyingness& \\
  & \code{Huber} & Huber's $\rho$ and $\psi$ functions& \\
  & \code{TukeyBisquare} & Tukey's $\rho$ and $\psi$ functions & \\
\cline{1-3}
  \multirow{5}{*}{Datasets} & \code{load\_telephone} & load telephone data& \multirow{5}{*}{} \\
  & \code{load\_stars} & load star cluster data & \\
  & \code{load\_animals} & load animals data  & \\
  & \code{load\_topgear} & load TopGear data  & \\
  & \code{load\_glass} & load glass data& \\
  \hline
\end{tabular}
\caption{Overview of the implemented methods per module.}
\label{tab:allmethods}
\end{table}
\end{center}
\normalsize

\subsection{Preprocessing} \label{sec:structure:preprocessing}

The data preprocessing module contains several classes. They are built using functionality from the base classes \code{BaseEstimator, OneToOneFeatureMixin} and \code{TransformerMixin}\linebreak from the \pkg{scikit-learn} API. This means they each implement a \code{.fit}, \code{.transform}, and\linebreak \code{.inverse\_transform} method.

The first class is \code{DataCleaner}, which operates similarly to the \code{checkDataSet} function from the \proglang{R}-package \pkg{cellWise}. The class is used to examine datasets and to exclude specific columns and rows that fail to meet certain conditions, e.g. by containing too many missing values. 
It works with a \code{fit} and a \code{transform} method, and it has two properties, \code{dropped\_columns} and \code{dropped\_rows}, to inspect which columns and rows were removed during the cleaning process. This preprocessor is mostly meant as a convenience tool that combines many conventional preprocessing steps into a single function.

A second class is \code{RobustPowerTransformer}, which robustly transforms skewed variables to approximate normality. For this purpose it employs a Yeo-Johnson or a Box-Cox transform as described in \cite{robustboxcoxyeojohnson}. The \code{fit} method robustly 
estimates a transformation parameter. The user can specify the type of transform (\code{"yeojohnson"} or \code{"boxcox"}). The default is the Yeo-Johnson transform which can handle negative data values. The transformation can then be applied using the \code{transform} method and can be inverted using \code{inverse\_transform}. The implementation is based on the function \code{transfo} in the \proglang{R}-package \pkg{cellWise} and can be considered as a robust alternative to \pkg{scikit-learn}'s \code{PowerTransformer}.

The third class is \code{RobustScaler}, constructed to scale features using a robust scale and/or location estimator of choice. By default the univariate Minimum Covariance Determinant described in the next section is applied, but the user can also pass a different \code{RobustScale}. This preprocessor is similar to \pkg{scikit-learn}'s \code{RobustScaler}, but is more flexible as it allows any robust scale estimator as opposed to only a quantile range.

\subsection{Univariate} \label{sec:structure:univariate}

The univariate module contains several univariate location and scale estimators from the robustness literature. They all use the implemented base class \code{RobustScale}, which has the properties \code{location} and \code{scale} and can be fitted using the \code{fit} method. Each child class is expected to implement a \code{.\_calculate} method where the attributes \code{scale\_} and \code{location\_} are set. Consider a univariate dataset 
$X = \{x_1,\dots,x_n\}$ of sample size $n$. A simple location
estimator is the median of the dataset $\med(X) = \med_i(x_i)$, and the scale can be estimated by the median absolute deviation given by $\mbox{MAD}(X) =
\med_i|x_i - \med(X)|$ (multiplied by 1.4833 if we want the MAD to be consistent for the standard deviation at normally distributed data). 
A minimalist child class that uses the median for location and the MAD for scale would look like this:
\begin{verbatim}
import numpy as np
from scipy.stats import median_abs_deviation

class MedianMAD(RobustScale):
    def _calculate(self, X):
        self.location_ = np.median(X)
        self.scale_ = median_abs_deviation(X)
\end{verbatim}

The univariate Minimum Covariance Determinant (MCD) 
estimator \citep{rousseeuw1987bookdata} looks for the $h$-subset (that is, 
a subset of $X$ containing $h$ observations) that has the smallest variance. It then estimates location by the average of that $h$-subset, and estimates scale from its standard deviation. The parameter $h$ must be at least $n/2$ and can be specified by the user as the argument \code{alpha}, which sets $h$ equal to \code{alpha} times~$n$. The algorithm employs a reweighting step and can apply a consistency correction for normal data. More details on the MCD can be found in the next section which describes it multivariate version. 

Also one-step M-estimators are available.
These location and scale estimators are quite 
robust and computationally efficient. Their implementation is based on the \proglang{R}-function 
\mbox{\code{estLocScale}} in the package \pkg{cellWise}. The one-step M-estimators start from initial estimates $(\hmu_0,\hsigma_0)$ of location and scale, e.g. the median and the MAD, and afterward perform one reweighting step:
\begin{equation*}
   \hmu_1(X)= \frac{\sum_i w_{1,i} x_i}{\sum_i w_{1,i}} 
   \quad  \text{and} \quad \hsigma_1(X) = 
   \sqrt{\frac{\sigma_0^2}{n \delta}\sum_i w_{2,i} r_i^2}\;.
\end{equation*}
The weights depend on two weight functions $W_1$ and $W_2$ by
\begin{equation*}
   w_{1,i} = W_1(r_i)\quad \text{and} \quad
   w_{2,i} = W_2(r_i) \quad \text{with} \quad
   r_i = \frac{x_i-\hmu_0}{\hsigma_0}\;.
\end{equation*}
The constant $\delta$ above ensures consistency for normally distributed data. 

Also the $Q_n$ scale estimator of \cite{Qn} is included.
It is defined as the first \textbf{Q}uartile of the distances between the points. More formally,
\begin{equation*}
    Q_n = 2.219\,\{|x_i-x_j|:i<j\}_{(k)} 
    \quad \text{with} \quad k = \binom{h}{2} 
    \quad \text{for} \quad 
    h=\left\lfloor\dfrac{n}{2}\right\rfloor + 1.
\end{equation*}
It has much better statistical efficiency than the MAD, and
is computed by the fast algorithm of \cite{fastQn}. 

Lastly, \pkg{RobPy} also contains the $\tau$-estimator from \cite{tau_and_OGK}, which is a special case of the one-step M-estimators given by
\begin{equation*}
    \tau_{location}= \frac{\sum_i w_i x_i}{\sum_i w_i} 
    \quad \text{and} \quad \tau_{scale} = 
    \sqrt{\frac{\MAD^2(X)}{n}\sum_i \rho_{c_2}
    \left(\frac{x_i - \tau_{location}}{\MAD(X)}\right)}
\end{equation*}
where the weights $w_i$ are defined as
\begin{equation*}
    w_i = W_{c_1}\left( \frac{x_i - \med(X)}{\MAD(X)}\right)
    \quad \text{with} \quad 
    W_{c}(u) = \left(1-\left(\frac{u}{c}\right)^2\right)^2 
    I(|u| \leq c)
\end{equation*}
and $\rho_{c}(u) = \min(c^2,u^2)$.
The default values are $c_1 = 4.5$ and $c_2 = 3$, but 
different values can be provided to the class \code{Tau}.

\subsubsection*{Adjusted boxplot}

The standard boxplot is a widely used tool to visualize univariate data. It is based on the median as well as the interquartile range 
$\IQR = Q_3 - Q_1$ in which $Q_1$ is the first quartile and $Q_3$ is the third. It displays the median inside the box that goes from $Q_1$ to $Q_3$\,, and flags points outside the interval $[Q_1 - 1.5\,\IQR\,,\,Q_3 + 1.5\,\IQR]$ as outliers. However, the reasoning behind this choice is restricted to data from a symmetric distribution. Therefore \cite{hubert2008adjustedboxplot} suggested the adjusted boxplot. This plot uses a different boundary to detect outliers, based on the medcouple (MC) of \cite{brys2004medcouple}, a robust skewness measure that is positive when the data has a longer tail on the right and negative when the tail is on the left. It flags points as outliers when they fall outside the interval \vskip-2mm
$$  [Q_1 - 1.5 e^{-4\MC} \IQR,Q_3 + 1.5 e^{3\MC} \IQR] 
    \quad \text{when} \quad \MC \geq 0$$
\vskip-2mm
and outside
\vskip-4mm
$$ [Q_1 - 1.5 e^{-3\MC} \IQR,Q_3 + 1.5 e^{4\MC} \IQR] 
    \quad \text{when} \quad \MC < 0.$$
This makes the adjusted boxplot more appropriate for asymmetric data. For symmetric data ($\MC=0$) the adjusted boxplot coincides with the classical boxplot. It is included in \pkg{RobPy} as \code{adjusted\_boxplot} and utilizes the \proglang{Python} module \pkg{statsmodels} \citep{seabold2010statsmodels} for the \code{medcouple} function. 

\subsection{Covariance} \label{sec:structure:covariance}

Various robust estimators of covariance matrices (also called `scatter matrices') have been proposed in the past decades, with different properties. The covariance module implements five frequently used scatter estimators. They all use the new base class \code{RobustCovariance} which builds on the \code{EmpiricalCovariance} class in \pkg{scikit-learn}. 

Now the dataset $X = (\bx_1,\dots,\bx_n)^{\top} \in \mathbb{R}^{n \times p}$ is multivariate, with $n$ cases and $p$ numerical variables, and the goal is to estimate a $p \times p$ scatter matrix $\bSigma$ and a central location $\bmu$ which is a $p$-variate point. One of the many uses of such estimates is to compute a statistical distance of each observation $\bx_i$ to the center $\bmu$ relative to the scatter matrix $\bSigma$. When the true $\bmu$ and $\bSigma$ are known, the statistical distance is defined as
\begin{equation*}
    d(\bx_i, \bmu, \bSigma) = \sqrt{\left(\bx_i - \bmu\right)^{\top} \bSigma^{-1} (\bx_i - \bmu)}\;.
\end{equation*}
The classical Mahalanobis distance is given by
\begin{equation*}
    \MD(\bx_i) = d(\bx_i, \boldsymbol{\overline{x}}, \cov(X))
\end{equation*}
where $\boldsymbol{\overline{x}}$ is the empirical mean and $\cov(X)$ is the empirical covariance matrix. The robust distances $\RD(\bx_i)$ are defined analogously, but they plug in a robust location for $\bmu$ and a robust scatter matrix for $\bSigma$.

The \code{RobustCovariance} class includes the distance-distance plot. It shows the robust distances versus the classical Mahalanobis distances, and is equipped with thresholds for outlier detection \citep{rousseeuw1999fastMCD}. By default, it uses the threshold $\sqrt{\chi_{p,0.975}^2}$ to flag points as outliers. The plot is drawn by the \code{distance\_distance\_plot} function, after obtaining a robust covariance estimator by the \code{fit} method.

\subsubsection*{Minimum Covariance Determinant}

The objective of the Minimum Covariance Determinant (MCD) estimator of \cite{rousseeuw1984LTSregression} is to find the $h$-subset whose empirical covariance matrix has the lowest determinant. More formally, we look for a subset $H \subset X$ with $\#H=h$ that minimizes $\text{det}(\cov(H))$, and then put the raw MCD estimates equal to
\begin{align*}
    \bmu_{MCD} =&\, \frac{1}{h}\sum_{\bx_i \in H} \bx_i\\
    \bSigma_{MCD} =&\, \frac{1}{h-1}\sum_{\bx_i \in H} (\bx_i - \bmu_{MCD})(\bx_i - \bmu_{MCD})^{\top}\;.
\end{align*}
The estimated scatter matrix $\bSigma_{MCD}$ is often multiplied by a correction factor to make it consistent when the data come from a normal distribution without outliers \citep{Pison:Corfac}. For other properties of the MCD see \cite{Cator2012} and \cite{MCD2018}. \cite{Cuesta2008} applied the MCD to cluster analysis.

Several algorithms exist for the MCD. The \pkg{RobPy} package implements two of them. The first one, \code{FastMCD}, carries out the algorithm of \cite{rousseeuw1999fastMCD}. It starts from a fixed number \code{n\_initial\_subsets} of random initial subsets. To each of them it applies so-called concentration steps (C-steps) that always lower the determinant, and it keeps the fit with the lowest determinant.
A second approach is implemented as \code{DetMCD}, based on the deterministic algorithm of \cite{hubert2012detMCD}. This algorithm is faster because it starts from only six carefully selected preliminary scatter estimators, each of them followed by C-steps until convergence. Both the above algorithms have an argument \code{alpha}, which sets $h$ equal to \code{alpha} times $n$. In practice, the raw MCD is almost always followed by a reweighting step to increase its efficiency. The weights for this are computed as
\begin{equation*}
    w_i = I\left(d(\bx_i, \bmu_{MCD}, \bSigma_{MCD}) \leqslant \sqrt{\chi_{p,0.975}^2}\right),
\end{equation*}
where ($\bmu_{MCD},\bSigma_{MCD}$) are the raw estimates. The reweighted MCD is then given by
\begin{align*}
    \bmu_{RMCD} &= \frac{1}{\sum_{i=1}^n w_i}\,\sum_{i=1}^nw_i \bx_i \\
    \bSigma_{RMCD}& = \frac{1}{\sum_{i=1}^n{w_i-1}}\,\sum_{i=1}^n w_i (\bx_i - \bmu_{RMCD})(\bx_i - \bmu_{RMCD})^{\top}\;.
\end{align*}
The boolean argument \code{reweighting} specifies whether the reweighting step is to be carried out. By default it is.

\subsubsection*{Cellwise MCD}

The algorithms described so far belong to the casewise framework, in which the word outlier refers to a case, that is, a row of the data matrix. But in recent years also a second framework is receiving attention. It can happen that most data cells
(entries) in a row are regular, and just a few of them are
anomalous. The first article to formulate the cellwise paradigm
was \cite{Alqallaf2009}. They noted how outliers propagate:
given a fraction $\varepsilon$ of contaminated cells at random positions, the expected fraction of contaminated rows is
$1 - (1 - \varepsilon)^p$ which grows quickly for increasing $\varepsilon$  and/or increasing dimension $p$. 
The two paradigms are quite different. The casewise 
paradigm is about cases that do not belong in the dataset, for
instance, because they are members of a different population. In contrast, the cellwise paradigm assumes that some cells in a row of the data matrix may deviate
from the values they should have had, perhaps due to measurement
errors, whereas the remaining cells in the same row
still contain useful information. See also \cite{ALYZ2015}.

Both of the above MCD algorithms fall within the casewise framework as they look for an optimal subset containing $h$ cases. However, within the cellwise framework another MCD estimator exists, namely the cellwise MCD estimator of \cite{raymaekers2023cellwiseMCD}. In this method $h$ no longer represents the number of inlying cases, instead it refers to the minimal number of unflagged cells per column (variable). It is implemented in the \pkg{RobPy} package as \code{CellMCD}, which mimics the functionality of the function \code{cellMCD} in the \proglang{R}-package \pkg{cellWise}. For this cellwise method, the usual distance-distance plot does not tell the whole story as it focuses on outlying cases. Instead, this method has its own \code{cell\_MCD\_plot} function that can make 5 different diagnostic plots depending on the argument \code{plottype}. The resulting plots will be illustrated in Section \ref{sec:cellwise_softwareusage}. An advantage of \code{CellMCD} is its built-in mechanism to handle missing values.

\subsubsection*{Wrapping covariance estimator}

As (ultra-) high-dimensional data becomes increasingly common today, \pkg{RobPy} also contains the wrapping covariance estimator of \cite{raymaekers2021froc}. This algorithm robustly standardizes the variables $x_{\bdot j}$ to new variables $z_{\bdot j}$ and then `wraps' them by applying the tailored data transformation $\psi$ given by
\begin{equation*}
    \psi_{b,c}(z) = \begin{cases}
  z  & \text{if} \  0 \leq |z| \leq b \\
  q_1 \tanh(q_2(c-|z|))\text{sign}(z) & \text{if} \ b \leq |z| \leq c \\
  0 & \text{if} \ c \leq |z|\;.
\end{cases}
\end{equation*}
Next it computes the classical covariance of the transformed data and undoes the standardization.
Default values of $b$ and $c$ are 1.5 and 4, with the corresponding $q_1$ and $q_2$ given in Appendix A.6 of \cite{raymaekers2021froc}. The wrapping approach is less sophisticated than the above algorithms, but on the other hand its simplicity allows fast computation of a fairly robust covariance estimator for high-dimensional data, which might have casewise as well as cellwise outliers. It is implemented as the function \code{WrappingCovariance}.

\subsubsection*{Other covariance estimators}

The covariance module contains two additional casewise covariance estimators. One of them is \code{KendallTau}, which estimates the covariance matrix by combining Kendall's $\tau$ rank correlation (implemented in \pkg{SciPy}) with univariate scales obtained by the \code{Qn} estimator, as in \cite{Ollerer2015}. A second covariance estimator is the Orthogonalized Gnanadesikan-Kettenring estimator proposed by \cite{tau_and_OGK}, which is one of the initial estimators used inside \code{DetMCD}.

\subsection{Regression} \label{sec:structure:regression}

Regression is a cornerstone of data science. It aims to predict a numerical response variable $y$ from one or more predictor variables $x_{\bdot j}$ that are also called features or regressors. Naturally, it has received considerable attention in robust statistics, which is why \pkg{RobPy} includes three of the most commonly used robust multiple linear regression algorithms. A base class is used, built on the \code{RegressorMixin} and the \code{BaseEstimator} classes of \pkg{scikit-learn}, employing a \code{fit} and a \code{predict} method. The regression base class is equipped with a tailored plotting tool called \code{outlier\_map}, which plots the standardized robust residuals versus the robust distances of the $x$-points formed by the predictor variables. The thresholds of the robust standardized residuals are $\pm 2.5$ and shown as horizontal lines, whereas the threshold on the robust $x$-distances is at $\sqrt{\chi_{p,0.975}^2}$ yielding a vertical line. This allows us to distinguish between\\ \vspace{-7mm}
\begin{itemize}
\setlength{\parskip}{-1pt}
\setlength{\itemsep}{2pt}
\item regular observations (inlying residual and inlying $x$-distance);
\item vertical outliers (outlying residual and inlying $x$-distance);
\item good leverage points (inlying residual and outlying $x$-distance); 
\item bad leverage points (outlying residual and outlying $x$-distance).
\end{itemize}  
\vspace{-3mm}
The plot is drawn by the function \code{outlier\_map} after fitting a robust regression.

\subsubsection*{Least Trimmed Squares}

A first robust regression estimator is the Least Trimmed Squares (LTS) estimator of \cite{rousseeuw1984LTSregression}. Similar to the MCD covariance estimator, it tries to find an $h$-subset with the smallest sum of squared residuals. This can also be formulated as
\begin{equation*}
    \hbbeta = \underset{\bbeta}{\text{argmin}} \sum_{i=1}^h (r(\bbeta)^2)_{(i)}
\end{equation*}
where $r_i(\bbeta)^2$ are the squared residuals given a vector $\bbeta$ of coefficients, which are then ordered as $(r(\bbeta)^2)_{(1)} \leq \dots \leq (r(\bbeta)^2)_{(n)}$\,. So the objective function is a sum of squared residuals, but it only adds $h$ of them instead of all. The size of this subset can be set by the argument \code{alpha}, which sets $h$ equal to \code{alpha} times $n$. The \code{FastLTSRegression} algorithm starts from \code{n\_initial\_subset} random initial subsets, applies C-steps to each of them, and keeps the result with the lowest objective. In order to increase the statistical efficiency of LTS it is usually followed by a reweighting step, which can be controlled by the boolean argument \code{reweighting}. 

\subsubsection*{S-estimators and MM-estimators}

The regression module also contains two other classes of robust regression estimators. An S-estimator \citep{Sest1984} is designed to minimize an M-estimator of scale of the residuals. It is given by
\begin{equation*}
  \hbbeta = \underset{\bbeta}{\text{argmin}} \ S(\bbeta)
  \quad \text{with} \quad \frac{1}{n} \sum_{i=1}^n \rho \left( \frac{r_i(\bbeta)}{S(\bbeta)} \right) = \frac{1}{2} 
\end{equation*}
where the function $\rho$ is specified by the argument \code{rho} that has a default (Tukey's bisquare) with good properties. 
An MM-estimator \citep{yohai1987MMregr} follows an S-estimator by a second step, similar to reweighting, to increase its statistical efficiency. The S-estimator is implemented as \code{SRegression} and uses the FAST-S algorithm of \cite{2006fastS}, and \code{MMRegression} combines FAST-S with iteratively reweighted least squares.

\newpage
\subsection{Principal component analysis} \label{sec:structure:PCA}

Principal component analysis (PCA) is a popular dimension reduction technique. Its classical version projects the $p$-variate datapoints on a $q$-dimensional subspace while retaining as much variance as possible. However, classical PCA is known to be very sensitive to outliers. Therefore, \pkg{RobPy} includes two robust PCA algorithms. They are built on the base class \code{RobustPCA}, which uses the \code{\_BasePCA} class from \pkg{scikit-learn}. The \code{fit} method learns the robust PCA components, after which the \code{transform} method obtains the $q$-variate scores and the \code{predict} method yields the projected points in $p$ dimensions. The new PCA module also provides a specialized plot \citep{hubert2005robpca} implemented as  \code{plot\_outlier\_map}. It displays the orthogonal distances, which are defined as the Euclidean distance from each point to its projection onto the PCA subspace, versus the score distances, i.e. the robust distances within the PCA subspace. This allows the user to distinguishes between regular cases, orthogonal outliers, good PCA leverage points, and bad PCA leverage points, in a way analogous to the outlier map of robust regression.  

\subsubsection*{ROBPCA}

The first robust PCA algorithm is \code{ROBPCA} from \cite{hubert2005robpca}. It combines the concept of projection pursuit with the idea of using eigenvectors of a robust covariance matrix of the scores. Its main steps are:\\ \vspace{-8mm}
\begin{enumerate}
\setlength{\parskip}{-1pt}
\setlength{\itemsep}{2pt}
\item Find the $h$ least outlying datapoints by selecting those with the lowest Stahel-Donoho outlyingness (to be described in Section \ref{sec:structure:utils});
\item Project the datapoints on the $q$-dimensional subspace spanned by the eigenvectors of the covariance matrix of these $h$ datapoints;
\item Carry out the reweighted MCD on the $q$-variate scores and use the eigendecomposition of its covariance matrix to obtain the final principal components.    
\end{enumerate}

\subsubsection*{Spherical PCA}

Spherical PCA \citep{locantore1999SPCA} is an earlier and less sophisticated PCA algorithm. It computes the principal components as the eigenvectors of the covariance matrix after projecting the data on a sphere, given by
\begin{equation}
    \frac{1}{n-1}\sum_{i=1}^n \frac{(\bx_i-\bhmu)(\bx_i-\bhmu)^{\top}}{||\bx_i-\bhmu||^2}
\end{equation}
where $\bhmu$ is a robust center such as the spatial median, which is the point $\bhmu$ with the lowest total distance
$\sum_{i=1}^n ||\bx_i - \bmu||$ from the datapoints. Spherical PCA is implemented as \code{PCALocantore}. As arguments one can give the desired number of principal components \code{n\_components}, or the fraction of the variance \code{k\_min\_var\_explained} that should be explained. 

\subsection{Detecting cellwise outliers} \label{sec:structure:outliers}

\pkg{RobPy} contains a recent cellwise outlier detection algorithm, the DetectDeviatingCells (DDC) algorithm of \cite{rousseeuw2018DDC}, implemented as the function \code{DDC}. This algorithm is suited even for scenarios with a high number of variables, and can also handle missing values. In addition to detecting outlying cells, it can predict them and impute missing values using the \code{fit}, \code{predict} and \code{impute} methods. \code{DDC} comes with a heatmap plotting tool called \code{cellmap}, used to visualize the standardized residuals of the DDC analysis. The implementation is based on the function \code{DDC} in the \proglang{R}-package \code{cellWise}.

\subsection{Utils} \label{sec:structure:utils}

Several of the methods and algorithms mentioned above utilize fundamental techniques. One of these is the Mahalanobis distance, discussed in Section \ref{sec:structure:covariance}. Another is the Stahel-Donoho outlyingness of a point $\bx$ relative to a dataset $(\bx_1,\dots,\bx_n )^{\top}$ of $p$ dimensions, defined as:
\begin{equation}
    \text{SDO}(\bx) = \sup_{\ba \in \mathbb{R}^p} \frac{|\ba^{\top}\bx - \med_i(\ba^{\top}\bx_i)|}{\MAD_i(\ba^{\top}\bx_i)}.
\end{equation}
Basic functions such as these are bundled in the utils module. For example, there is a base class \code{BaseRho} containing the $\rho$- and $\psi$-functions \code{TukeyBisquare} and \code{Huber}, so they can easily be used. The functions \code{mahalanobis\_distance} and \code{stahel\_donoho} provide straightforward implementations of the above building blocks. The univariate weighted median implemented as \code{weighted\_median} is used in the computation of the $Q_n$ estimator, and the spatial median given by \code{l1median} is used in spherical PCA. 

\subsection{Datasets} \label{sec:structure:datasets}

\pkg{RobPy} includes several datasets that are often encountered in the robustness literature. These datasets serve as standard examples and benchmarks, allowing users to easily test robust algorithms. They are listed in Table \ref{tab:datasets}. The first three datasets consist of bivariate data with small sample sizes, and are commonly employed to illustrate robust covariance and regression estimators. They are also available in the \proglang{R}-package \pkg{robustbase}. The fourth is the well-known TopGear dataset originating from the \proglang{R}-package \pkg{robustHD}. It contains 32 variables of mixed types about 297 cars that were featured in the BBC television show Top Gear until 2014. The TopGear data has been utilized in different settings, including PCA, regression, and cellwise outlier detection. The final dataset is the high-dimensional glass data, consisting of 180 observed intensities at 750 wavelengths. This spectral dataset has frequently been used to illustrate casewise and cellwise outlier detection and is also available in the \proglang{R}-package \pkg{cellWise}.

\begin{table}[H]
    \centering
    \begin{tabular}{|c|c|c|c|c|c|}
    \hline
    data & name in \proglang{R}-package & \proglang{R}-package & n & p & reference\\
    \hline
    telephone & telef & robustbase & 24 & 2 & \cite{rousseeuw1987bookdata}\\
    stars & starsCYG & robustbase & 47 & 2 & \cite{rousseeuw1987bookdata}\\
    animals & Animals2 & robustbase & 65 & 2 & \cite{rousseeuw1987bookdata}\\
    topgear & TopGear & robustHD & 297 & 32 & \cite{robustHD}\\
    glass & data\_glass & cellWise & 180 & 750 & \cite{lemberge2000glassdata}\\
    \hline
    \end{tabular}
    \caption{The datasets available in \pkg{RobPy}.}
    \label{tab:datasets}
\end{table}

The datasets can be loaded using the functions \code{load\_telephone}, \code{load\_stars}, \code{load\_animals}, \code{load\_topgear} and \code{load\_glass}. This returns a \code{Bunch} object from \pkg{scikit-learn}, containing the data, a list of the names of the features, a description of the dataset and a path to the location of the data.

\section{Software usage} \label{sec:softwareusage}

In this section, we demonstrate the implemented robust algorithms by applying them to the TopGear data. This dataset is available in \pkg{RobPy} in the module \code{datasets} and can be loaded as follows:

\begin{verbatim}
>>> from robpy.datasets import load_topgear
>>> data = load_topgear(as_frame=True)
>>> print(data.DESCR)
>>> data.data.head()
\end{verbatim}

These commands also show the corresponding data description and the first 5 rows.

Prior to showcasing the methods, we remark that notebooks that execute all code in this section are available at \href{https://robpy.readthedocs.io/en/latest/examples.html}{robpy.readthedocs.io}\;.

\subsection{Preprocessing} \label{sec:preprocessing_softwareusage}

\begin{figure}[H]
    \centering
    \includegraphics[width=0.34\linewidth]{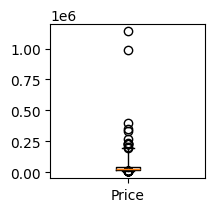}
    \includegraphics[width=0.5\linewidth]{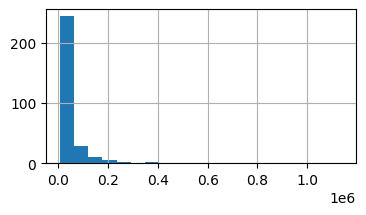}
    \captionsetup{skip=0pt}
    \caption{Adjusted boxplot and histogram of the variable \code{Price}.}
    \label{fig:price}
\end{figure}

Before we subject the TopGear data to various methods we apply some preprocessing steps, as is often done when analyzing data. First we clean the data by removing columns and rows that are not suited for most analysis methods, such as non-numerical columns, columns and rows with too many missing values, discrete columns, columns with a scale of zero, and columns corresponding to the case numbers. We do this with the constructed class \code{DataCleaner}:
\begin{verbatim}
>>> from robpy.preprocessing import DataCleaner
>>> cleaner = DataCleaner().fit(data.data)
>>> clean_data = cleaner.transform(data.data)
\end{verbatim} 
To see which columns and rows have been removed, we run the following code: 
\begin{verbatim}
>>> print(json.dumps(cleaner.dropped_columns, indent=4))
>>> print(cleaner.dropped_rows)
\end{verbatim}
This gives us:
\begin{verbatim} 
{
    "non_numeric_cols": [
        "Make",
        "Model",
        "Type",
        ...
        "Origin"
    ],
    "cols_rownumbers": [],
    "cols_discrete": [
        "Fuel",
        "DriveWheel",
        "AdaptiveHeadlights",
        ...
        "Origin"
    ],
    "cols_bad_scale": [
        "Cylinders"
    ],
    "cols_missings": []
}
{'rows_missings': [69, 95]}
\end{verbatim} 
As a second preprocessing step, we aim to robustly transform variables that are skewed and/or have a long tail, to bring them closer to normality. For example, the variable \code{Price} is clearly right skewed, as we can see in its adjusted boxplot and its histogram in Figure~\ref{fig:price}.
\begin{verbatim}
>>> adjusted_boxplot(clean_data['Price'],figsize=(2,2))
>>> clean_data['Price'].hist(bins=20, figsize=(4, 2))
\end{verbatim}

\begin{figure}[H]
    \centering
    \includegraphics[width=0.286\linewidth]{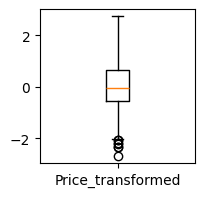}
    \includegraphics[width=0.5\linewidth]{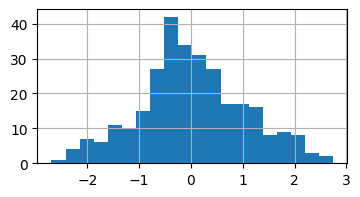}
    \caption{Adjusted boxplot and histogram of the variable \code{Price} transformed by the Box-Cox transformation with $\lambda = -0.42354$\,.}
    \label{fig:pricetransformed}
\end{figure}

\begin{figure}[H]
    \centering
    \includegraphics[width=1\linewidth]{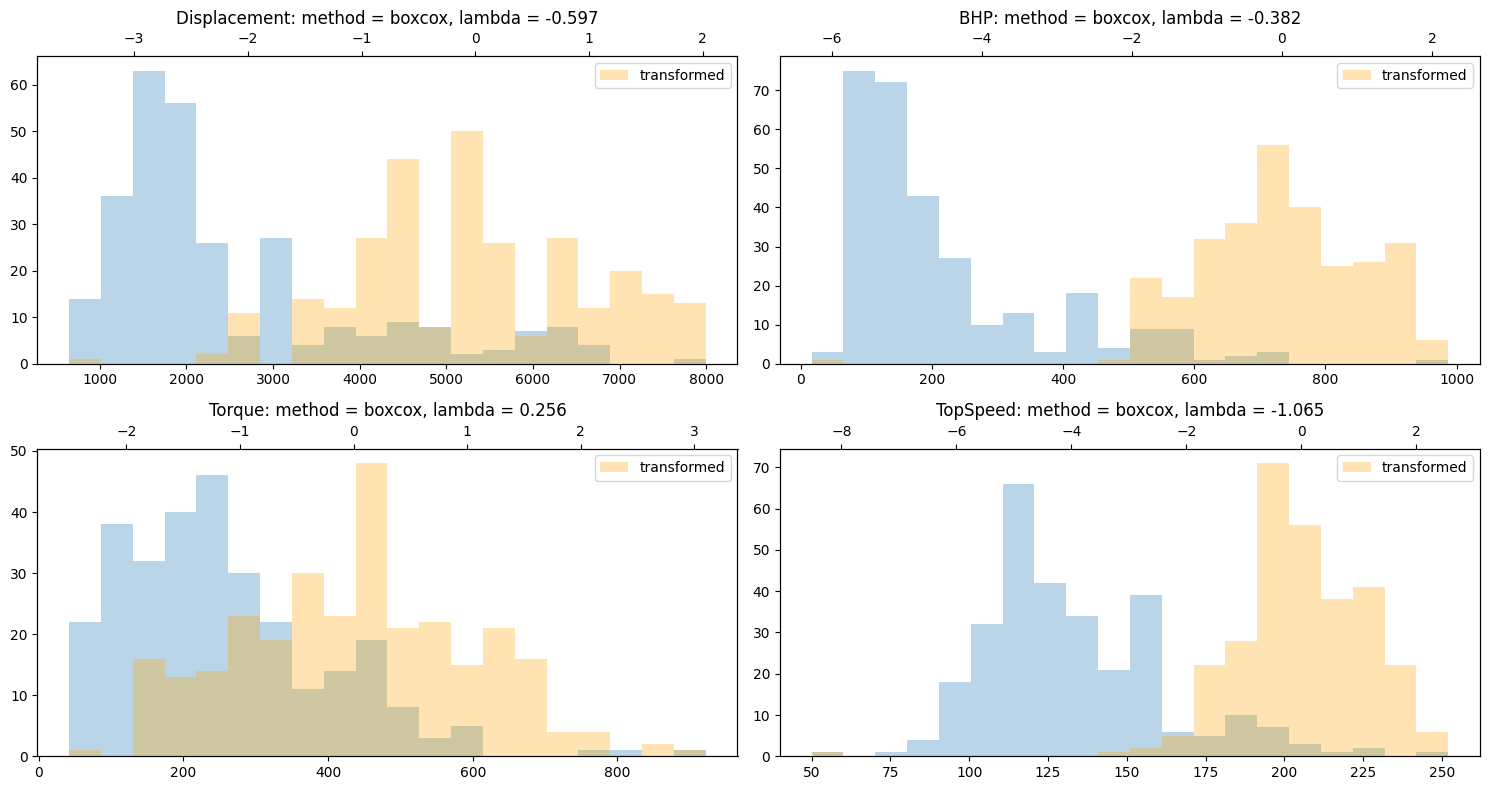}
    \caption{\code{RobustPowerTransformer} applied to
    \code{Displacement, BHP, Torque} and \code{TopSpeed}.}
    \label{fig:othertransforms}
\end{figure}

As this might affect approaches that assume normality or symmetry of the inliers, we now use the class \code{RobustPowerTransformer} to robustly transform the variable so its central part becomes more symmetric:
\begin{verbatim}
>>> from robpy.preprocessing import RobustPowerTransformer
>>> price_transformer = RobustPowerTransformer(method='auto').fit
        (clean_data['Price'])
>>> clean_data['Price_transformed'] = price_transformer.transform
        (clean_data['Price'])
\end{verbatim}
The \code{fit} method of \code{RobustPowerTransformer} selects the most appropriate transformation, which is either of the Yeo-Johnson type or the Box-Cox type. It estimates the transformation parameter $\lambda$ by minimizing a reweighted maximum likelihood objective function. Afterward the \code{transform} method applies the transformation to the data variable. For \code{Price} the selected transformation and $\lambda$ parameter are:
\begin{verbatim}
>>> price_transformer.method, price_transformer.lambda_rew
\end{verbatim}
\begin{verbatim}
('boxcox', -0.42354039562300644)
\end{verbatim}
The method has selected the Box-Cox transformation here. A parameter value of $\lambda=1$ would indicate no transformation, and $\lambda=0$ would correspond to the logarithmic transform. The selected $\lambda \approx -0.42$ thus goes a bit further than the logarithmic transform.
Figure~\ref{fig:pricetransformed} shows the robustly transformed variable. We see that the transformation has made the variable more symmetric and has reduced the long right tail.
\begin{verbatim}
>>> adjusted_boxplot(clean_data['Price_transformed'],figsize=(2,2))
>>> clean_data['Price_transformed'].hist(bins=20, figsize=(4, 2))
\end{verbatim}
Other skewed variables that benefit from a robust power transformation are \code{Displacement, BHP, Torque} and \code{TopSpeed}:
\begin{verbatim}
>>> fig, axs = plt.subplots(2, 2, figsize=(15, 8))
>>> for col, ax in zip(['Displacement', 'BHP', 'Torque', 'TopSpeed'],
        axs.flatten()):
>>>     clean_data[col].hist(ax=ax, bins=20, alpha=0.3)
>>>     transformer = RobustPowerTransformer(method='auto')
            .fit(clean_data[col].dropna())
>>>     clean_data.loc[~np.isnan(clean_data[col]), col] = 
            transformer.transform(clean_data[col].dropna())
>>>     ax2=ax.twiny()
>>>     clean_data[col].hist(ax=ax2, bins=20, label='transformed',
            color='orange', alpha=0.3)
>>>     ax.grid(False)
>>>     ax2.grid(False)
>>>     ax2.legend(loc='upper right')
>>>     ax.set_title(f'{col}: method = {transformer.method}, lambda = 
            {transformer.lambda_rew:.3f}')
>>> fig.tight_layout()
\end{verbatim}
The transformation results for these variables are shown in Figure~\ref{fig:othertransforms}. In each panel the blue histogram shows the original variable, the orange histogram shows the transformed variable, and the overlap has a mixed color.
Another preprocessing class in the \pkg{RobPy} package is the \code{RobustScaler} class, designed to scale and/or center variables using a \code{RobustScale}. This could for instance be the location and the scale from the univariate Minimum Covariance Determinant, implemented as \code{UnivariateMCD}. We will illustrate it in Section \ref{sec:PCA_softwareusage}.

\subsection{Location and scatter estimators} \label{sec:locationscatter_softwareusage}

Estimates of location and scatter play a key role in data analysis. It is widely known that the classical mean and covariance matrix are heavily influenced by outliers. Therefore we implemented several robust alternatives, all building on the base class \code{RobustCovariance}. There are two algorithms for the Minimum Covariance Determinant, \code{FastMCD} and \code{DetMCD}. The Orthogonalized Gnanadesikan-Kettenring estimator is computed by \code{OGK}, Kendall's tau covariance by \code{KendallTau}, and the covariance matrix based on the wrapping function is obtained by \code{WrappingCovariance}. Below we illustrate the functionality of \code{FastMCD} on the TopGear data, but the other methods are run in a similar way.

As the estimators listed above only work on data without missing values, we first remove the cars containing missing values. However, in Section \ref{sec:cellwise_softwareusage} a cellwise robust covariance algorithm will be illustrated that can handle missing values. 
\begin{verbatim}
>>> clean_data2 = clean_data.dropna()
\end{verbatim}
Next, we calculate the Minimum Covariance Determinant on the data by employing the \code{fit} method. Here we use the transformed \code{Price} variable, so we remove the original \code{Price} variable. To visualize the outliers we draw a distance-distance plot. 
\begin{verbatim}
>>> from robpy.covariance import FastMCD
>>> mcd = FastMCD().fit(clean_data2.drop(columns=['Price']))
>>> fig = mcd.distance_distance_plot()
\end{verbatim}
\begin{figure}[H]
    \centering
    \includegraphics[width=.5\linewidth]{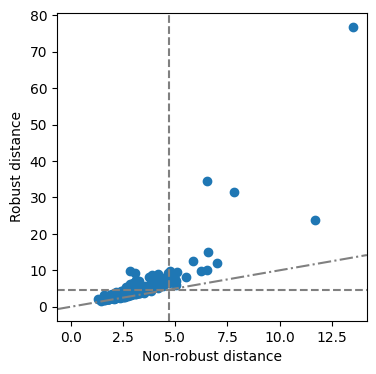}
    \caption{Distance-distance plot showing the robust distances obtained by the fast MCD estimator versus the classical Mahalanobis distances (non-robust distances).}
    \label{fig:fastmcd}
\end{figure}
In Figure~\ref{fig:fastmcd}, one extreme outlier stands out in the top right corner, as it has a very large robust distance as well as a large Mahalanobis distance. We can inspect this outlier as follows:
\begin{verbatim}
>>> data.data.loc[
       clean_data2.index[(mcd._robust_distances > 60) & 
       (mcd._mahalanobis_distances > 12)], ['Make', 'Model']+
       list(set(clean_data2.columns).intersection(set(data.data.columns)))]
\end{verbatim}
\begin{verbatim}
      Make   Model  Acceleration  Height  ...  Price    Displacement
41    BMW    i3     7.9           1578.0  ...  33830.0  647.0
\end{verbatim}
The outlier is the electric BMW i3, which stands out because this is a dataset featuring vehicles from before 2014, when there were few electric cars.

\subsection{Principal component analysis} \label{sec:PCA_softwareusage}

As principal component analysis is usually based on the classical covariance estimator, it is very prone to outliers. In the \pkg{RobPy} package, we have implemented two robust PCA methods: \code{ROBPCA} and spherical PCA (\code{PCALocantore}). Here we will illustrate \code{ROBPCA}.

\begin{figure}[H]
    \centering
    \includegraphics[width=.5\linewidth]{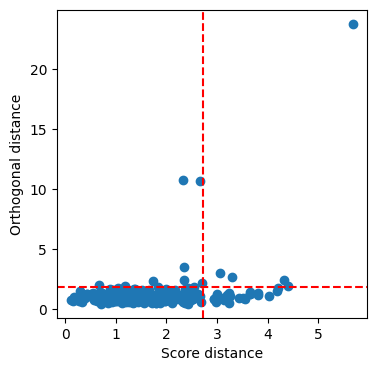}
    \caption{PCA outlier map for the ROBPCA analysis.}
    \label{fig:robpca}
\end{figure}

When the variables have different measurement units or some variables have very different scales, any PCA analysis will be dominated by the variables with the largest scale. Therefore, it is often recommended to scale the variables first. Here we start by scaling the TopGear data with the \code{RobustScaler} class, using its default scale estimator, which is the univariate MCD for each variable:
\begin{verbatim}
>>> from robpy.preprocessing import RobustScaler
>>> from robpy.pca import ROBPCA
>>> scaled_data = RobustScaler(with_centering=False).fit_transform(
        clean_data2.drop(columns=['Price']))
>>> pca = ROBPCA().fit(scaled_data)
>>> print(pca.components_)
>>> print(pca.explained_variance_ratio)
\end{verbatim}
\begin{verbatim}
[[-0.3249863   0.09160325]
 [-0.32181544  0.17999333]
 [-0.33824779 -0.00529083]
 [ 0.25809292 -0.34035504]
 [-0.24901036  0.31572018]
 [ 0.2507995  -0.14444022]
 [-0.36398309 -0.30642778]
 [-0.35399797 -0.24551552]
 [-0.34291378 -0.2203963 ]
 [-0.06770964 -0.71357778]
 [-0.32390731  0.10419143]]
[0.75655623 0.87247218]
\end{verbatim}
The output indicates that ROBPCA has obtained 2 principal components that together  explain 87.25\% of the variance. If more components are wanted, this can be specified by the argument \code{n\_components}. To visualize possible outliers we draw an outlier map, shown in Figure~\ref{fig:robpca}.
\begin{verbatim}
>>> score_distances, orthogonal_distances, score_cutoff, od_cutoff = 
        pca.plot_outlier_map(scaled_data, return_distances=True)
\end{verbatim}
Most points in Figure~\ref{fig:robpca} lie in the region of regular cases, plus a few orthogonal outliers and good leverage points. In the top right panel we see an extreme bad leverage point, that is, a point with an outlying orthogonal distance and an outlying score distance, as well as some borderline cases. We identify these cars as follows:
\begin{verbatim}
>>> data.data.loc[clean_data2.loc[(score_distances > score_cutoff) &
        (orthogonal_distances > od_cutoff)].index, ['Make', 'Model'] + 
        list(set(clean_data2.columns).intersection(set(data.data.columns)))]
\end{verbatim}
\begin{verbatim}
    Make           Model    Displacement  TopSpeed  ...  MPG
41  BMW            i3       647.0         93.0      ...  470.0
49  Bugatti        Veyron   7993.0        252.0     ...  10.0
124 Jeep           Wrangler 2777.0        107.0     ...  34.0
135 Land Rover     Defender 2198.0        90.0      ...  25.0
164 Mercedes-Benz  G-Class  2987.0        108.0     ...  25.0
196 Pagani         Huayra   5980.0        230.0     ...  23.0
\end{verbatim}
Here we again observe an electric car (the BMW i3), two ultra-high-performance sports cars (Bugatti Veyron and Pagani Huayra) and three rugged all terrain vehicles (Jeep Wrangler, Land Rover Defender and the Mercedes-Benz G-Class). It makes sense that these stand out.

\subsection{Regression} \label{sec:regression_softwareusage}
We saw that \pkg{RobPy} contains several robust regression algorithms. Here we demonstrate the workings of the \code{MMRegression} class. The algorithms \code{FastLTSRegression} and \code{SRegression} operate similarly. 

\begin{figure}[H]
    \centering
    \includegraphics[width=.5\linewidth]{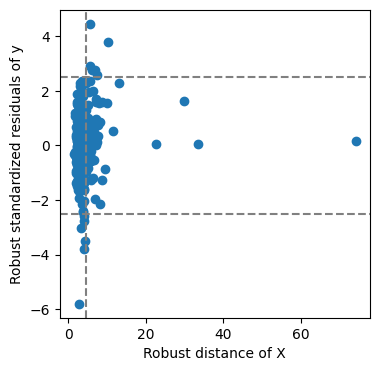}
    \caption{TopGear data: outlier map of the MM-regression of the transformed price on the other numerical variables.}
    \label{fig:MMregression}
\end{figure}

Our goal is to predict the transformed price from the other variables. The linear model is fitted as follows, resulting in the coefficients below:
\begin{verbatim}
>>> from robpy.regression import MMRegression
>>> X = clean_data2.drop(columns=['Price', 'Price_transformed'])
>>> y = clean_data2['Price_transformed']
>>> estimator = MMRegression().fit(X, y)
>>> estimator.model.coef_
\end{verbatim}
\begin{verbatim}
array([ 2.71338919e-01,  4.24224370e-01,  2.04835077e-01,  3.63925688e-02,
        8.75191584e-02,  3.77330897e-03,  4.65624836e-04, -2.97257613e-04,
        8.37238866e-04, -9.42993337e-04])
\end{verbatim}
Next we draw an outlier map to see whether there are any outliers:
\begin{verbatim}
>>> resid, std_resid, distances, vt, ht = estimator.outlier_map(X, 
        y.to_numpy(), return_data=True)
>>> bad_leverage_idx = (np.abs(std_resid) > vt) & (distances > ht)
\end{verbatim}
The resulting plot is Figure~\ref{fig:MMregression}. It considers most points as regular cases and good leverage points, plus a few vertical outliers. We also see some points that fall in the `bad leverage point' region, most of which are merely borderline cases.
As these points have an outlying price, we would like to compare the actual price with the price predicted by the model. However, in section \ref{sec:preprocessing_softwareusage} we transformed the variable \code{Price} by a Box-Cox transform, so the predictions need to be transformed back first:
\begin{verbatim}
>>> data.data.loc[clean_data2[bad_leverage_idx].index, ['Make', 'Model',
        'Price']].assign(predicted_price=price_transformer.inverse_transform
        (estimator.predict(X.loc[bad_leverage_idx])).round())
\end{verbatim}
\begin{verbatim}
        Make           Model          Price     predicted_price
2       Aston Martin   Cygnet         30995.0   15326.0
5       Aston Martin   V12 Zagato     396000.0  117938.0
164     Mercedes-Benz  G-Class        82945.0   34576.0
222     Rolls-Royce    Phantom        352720.0  116166.0
223     Rolls-Royce    Phantom Coupe  333130.0  111003.0
253     Toyota         Prius          24045.0   16272.0
\end{verbatim}
We see that these points represent cars from prestigious brands (Aston Martin, Mercedes and Rolls-Royce) and a pioneer in hybrid technology (the Toyota Prius). These characteristics explain why their actual price exceeds their predicted price.

\subsection{Algorithms for cellwise outliers} \label{sec:cellwise_softwareusage}
All algorithms illustrated so far are designed for the casewise outlier paradigm. We now switch to the cellwise outlier paradigm and illustrate the workings of the DetectDeviatingCells and cellMCD algorithms, both of which can also handle missing values.

\subsubsection*{DetectDeviatingCells} \label{sec:ddc_softwareusage}
The DetectDeviatingCells (DDC) algorithm \citep{rousseeuw2018DDC} takes a dataset that may contain missing values, and aims to find cellwise outliers. It computes robust correlations between the variables, and produces a predicted value for each cell based on the other cells in the same row. Each cell thus obtains a cellwise residual, which is the actual cell value minus its prediction. When the absolute value of the standardized cellwise residual is above $\sqrt{\chi^2_{1,0.99}} \approx 2.5$ the cell is flagged as outlying.

The results of DDC can be visualized in a cellmap. This is a kind of heatmap in which data cells are represented by small squares. Cells with inlying residuals are colored yellow, and missing values are white. Cells with a positive outlying residual are shown in red. These are measurements that were higher than predicted.
Cells with negative outlying residual are shown in blue instead. In order to distinguish far outliers from borderline cells, the color changes gradually from yellow over orange to more intense red, and from yellow over light purple to more intense blue.

The code below runs the DDC algorithm and draws such a cellmap. Since the TopGear data contains a lot of cars the full cellmap is very big, and mostly yellow because the percentage of outlying cells is small. For illustration purposes we only show the cellmap for a small subset of 17 cars, some of which have interesting outlying cells.
\begin{verbatim}
>>> from robpy.outliers import DDC
>>> ddc = DDC().fit(clean_data.drop(columns=['Price']))
>>> row_indices = np.array([ 11,  41,  55,  73,  81,  94,  99, 135, 150, 164,
        176, 198, 209, 215, 234, 241, 277])
>>> ax = ddc.cellmap(clean_data.drop(columns=['Price']), 
        row_zoom=row_indices)
>>> cars = data.data.apply(lambda row: f"{row['Make']} {row['Model']}", 
        axis=1).tolist()
>>> ax.set_yticklabels([cars[i] for i in row_indices], rotation=0)
\end{verbatim}
\begin{figure}[!ht]
    \centering
    \includegraphics[width=0.85\linewidth]{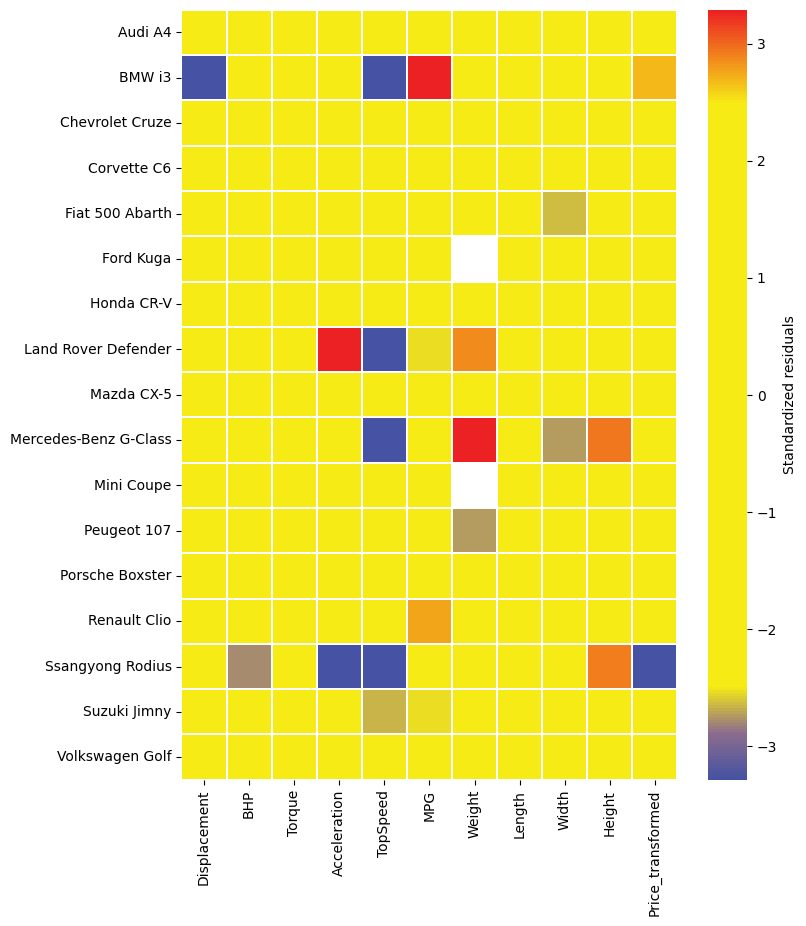}
    \caption{TopGear cellmap resulting from the DetectDeviatingCells algorithm.}
    \label{fig:DDCcellmap}
\end{figure}

The resulting cellmap is in Figure~\ref{fig:DDCcellmap}.  
We see for instance that the Audi A4 has no outlying cells, whereas the electric BMW i3 has an abnormally small \code{Displacement} (which is that of its small `range extender' petrol engine that can generate extra electricity) and very high miles per gallon \code{MPG}. Some of the other outlying cells in this cellmap are interpreted in \cite{rousseeuw2018DDC}.

Note that the \code{impute} method from the class \code{DDC} provides a version of the dataset in which the missing values and the outlying cells have been imputed, while leaving the remaining cells unchanged. This imputed dataset can be useful with other analysis methods. Moreover, when we receive new (out of sample) data, even if it is just a single case, the \code{predict} method will return the same type of results as for the training data, such as standardized cellwise residuals, detection of outlying cells, and cell imputation.

\subsubsection*{Cellwise MCD} \label{sec:cellmcd}
In this last section we demonstrate the cellwise MCD algorithm , a robust covariance estimator for the cellwise outlier paradigm. It is implemented as the class \code{CellMCD} and can also handle missing values. Here we apply it to the TopGear data. To reproduce the results in \cite{raymaekers2023cellwiseMCD} we first log-transform the variables \code{Displacement}, \code{BHP}, \code{Torque}, and \code{TopSpeed}.
\begin{verbatim}
>>> from robpy.covariance.cellmcd import CellMCD

>>> data = load_topgear(as_frame=True)
>>> car_models = data.data['Make'] + data.data['Model']
>>> cleaner = DataCleaner().fit(data.data)
>>> clean_data = cleaner.transform(data.data)
>>> clean_data = clean_data.drop(columns=['Verdict'])
>>> for col in ['Displacement', 'BHP', 'Torque', 'TopSpeed']:
>>>     clean_data[col] = np.log(clean_data[col])
>>> clean_data['Price'] = np.log(clean_data['Price']/1000)
>>> car_models.drop(cleaner.dropped_rows["rows_missings"],inplace=True)
>>> car_models = car_models.tolist()
>>> clean_data.head()

>>> cellmcd = CellMCD()
>>> cellmcd.fit(clean_data.values)
\end{verbatim}

\begin{figure}[H]
    \includegraphics[width=.48\textwidth]{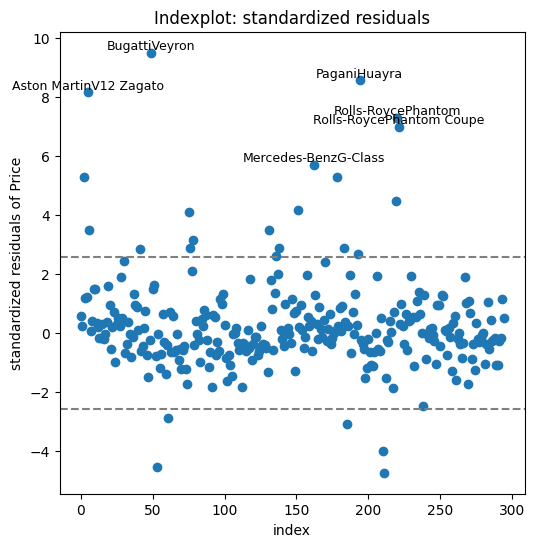}\hfill
    \includegraphics[width=.5\textwidth]{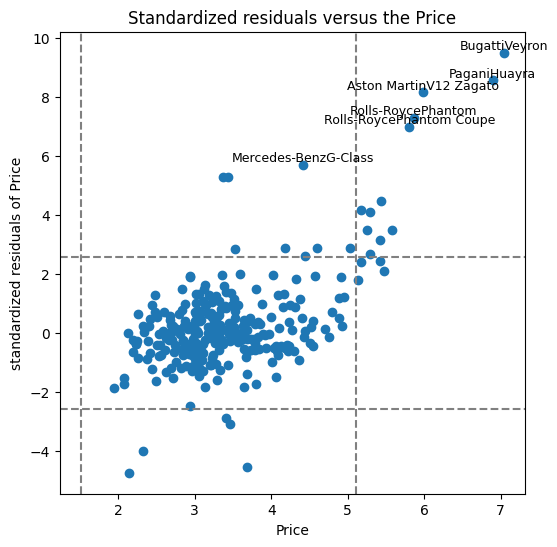}\hfill
    \\[\smallskipamount]
    \includegraphics[width=.5\textwidth]{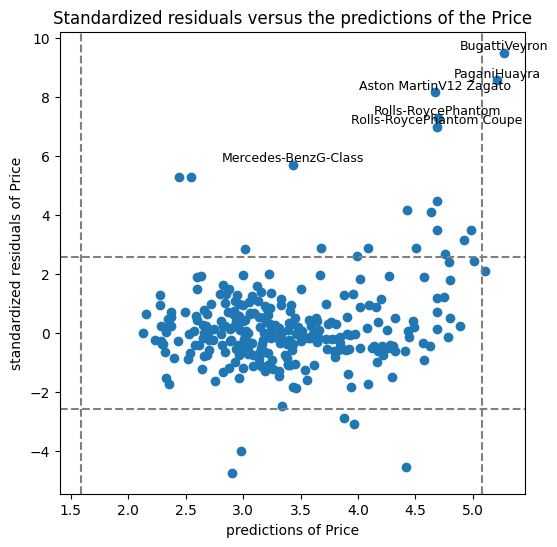}\hfill
    \includegraphics[width=.49\textwidth]{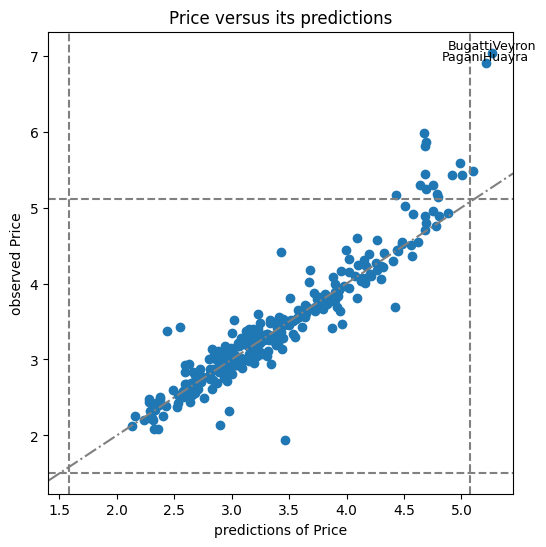}\hfill
    \caption{TopGear data: diagnostic plots from \code{CellMCD}.}
    \label{fig:cellmcdprice}
\end{figure}

We now illustrate the diagnostic plots available for cellMCD, which 
are geared towards identifying outlying cells instead of outlying cases. The four plots in Figure~\ref{fig:cellmcdprice} focus on a single variable, the log-transformed \code{Price}. The panels of Figure~\ref{fig:cellmcdprice} were obtained by the following commands, in which the log(Price) variable has index 0:
\begin{verbatim}
>>> cellmcd.cell_MCD_plot(variable=0, variable_name="Price", 
       row_names=car_models, plottype="indexplot", 
       annotation_quantile=0.9999999)
>>> cellmcd.cell_MCD_plot(variable=0, variable_name="Price", 
       row_names=car_models, plottype="residuals_vs_variable",
       annotation_quantile=0.9999999)
>>> cellmcd.cell_MCD_plot(variable=0, variable_name="Price", 
       row_names=car_models, plottype="residuals_vs_predictions",
       annotation_quantile=0.9999999)
>>> cellmcd.cell_MCD_plot(variable=0, variable_name="Price", 
       row_names=car_models, plottype="variable_vs_predictions",
       annotation_quantile=0.99999)
\end{verbatim}

The automatically annotated cars in Figure 8 were already flagged in Sections \ref{sec:PCA_softwareusage} and \ref{sec:regression_softwareusage}. They are high-end models from prestigious brands. Their price is not in line with their other characteristics according to the patterns formed by most of the cars. The price of these cars includes an exclusivity premium, as is typical for luxury goods.

\begin{figure}
    \centering
    \includegraphics[width=0.6\linewidth]{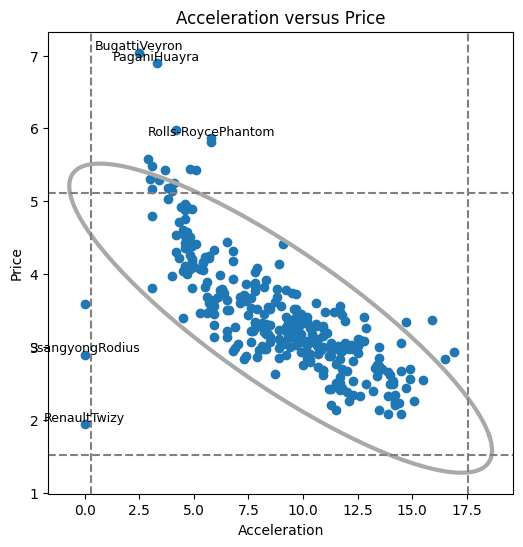}
    \caption{TopGear data: a bivariate diagnostic plot from \code{CellMCD} with tolerance ellipse.}
    \label{fig:cellmcdbivariate}
\end{figure}

We can also look at a scatterplot of two variables with their tolerance ellipse based on the cellwise MCD. Here we do this for \code{Acceleration}, which is the time needed  to accelerate from zero to 60 miles per hour, versus the logarithm of \code{Price}:

\begin{verbatim}
>>> cellmcd.cell_MCD_plot(second_variable = 4, 
       second_variable_name = "Acceleration", row_names = car_models, 
       variable = 0, variable_name = "Price", plottype = "bivariate", 
       annotation_quantile=0.999999)
\end{verbatim}

The resulting plot is Figure~\ref{fig:cellmcdbivariate}. We see a negative relation between acceleration time and price, and indeed the latter can be seen as a proxy for horsepower. The automatically annotated cars at the top were flagged before, but also two new outliers are detected, the SsangYong Rodius and the Renault Twizy. In the plot we see that these cars (and one other car) have a cell value of zero for \code{Acceleration}, which is physically impossible. The robust analysis has thus detected errors in the data. Presumably the true \code{Acceleration} values of these cars were not measured, and someone filled in zeroes instead of labeling these cells as NA.

\section*{Coding details}

The \pkg{RobPy} package currently uses the following versions of \proglang{Python} and accompanying \mbox{libraries}:
\begin{itemize}
    \item \proglang{Python} 3.10 as the programming environment;
    \item \pkg{scikit-learn} (v1.3 or higher) for machine learning algorithms and data processing;
    \item \pkg{SciPy} (v1.11.4) for scientific computing, including statistical functions;
    \item \pkg{statsmodels} (v0.14.1) for advanced statistical modeling;
    \item \pkg{Matplotlib} (v3.8.2) for data visualization and plotting \citep{matplotlib};
    \item \pkg{tqdm} (v4.66.1) for progress bar integration during computational tasks \citep{tqdm};
    \item \pkg{seaborn} (v0.13.2) for additional visualization features \citep{seaborn}.
\end{itemize}

\section*{Acknowledgments}

Sarah Leyder is supported by Fonds Wetenschappelijk onderzoek - Vlaanderen (FWO) as a PhD fellow Fundamental Research (PhD fellowship 11K5525N). This research received funding from the Flemish Government under the "Onderzoeksprogramma Artificiële Intelligentie (AI) Vlaanderen” programme.


\end{document}